\documentclass[superscriptaddress,showpacs,aps,prc,preprint,onecolumn]{revtex4-1}
\usepackage{dcolumn}
\usepackage{amsmath}
\usepackage{amssymb}
\usepackage[dvips]{epsfig}
\usepackage{enumerate}
\usepackage{bm}
\usepackage{graphicx}
\usepackage{units}
\usepackage{longtable}

\newcommand{\jp}{J^\pi}

\newcommand{\kp}{K^\pi}
\newcommand{\pp}{(\mathrm{p},\mathrm{p'})}
\newcommand{\ppg}{(\mathrm{p},\mathrm{p'}\gamma)}
\newcommand{\nucl}[2]{^{#1}\mathrm{#2}}
\newcommand{\figref}[1]{Fig.~\ref{#1}}
\newcommand{\tabref}[1]{table \ref{#1}}
\newcommand{\secref}[1]{section \ref{#1}}

\newcommand{\yale}{\affiliation{Wright Nuclear Structure Laboratory, Yale University, New Haven, CT 06520-8120, USA}}
\newcommand{\ikp}{\affiliation{Institut f\"ur Kernphysik, Universit\"at zu K\"oln, D-50937 K\"oln, Germany}}
\newcommand{\emmi}{\affiliation{ExtreMe Matter Institute EMMI and Research Division, GSI, Helmholtzzentrum, Darmstadt, Germany}}
\newcommand{\fias}{\affiliation{Frankfurt Institute for Advanced Studies FIAS, Frankfurt, Germany}}
\newcommand{\chalmers}{\affiliation{Fundamental Physics, Chalmers University of Technology, SE-412-916 G\"oteborg, Sweden}}
\newcommand{\rich}{\affiliation{University of Richmond, Richmond, VA, 23173, USA}}
\newcommand{\surrey}{\affiliation{Department of Nuclear Physics, University of Surrey, Guildford GU2 7XH, UK}}
\newcommand{\magu}{\affiliation{National Institute for Physics and Nuclear Engineering, R-77125, Bucharest-Magurele, Romania}}

\begin{document}

\title{Investigation of octupole vibrational states in $^{150}\mathrm{Nd}$ via inelastic proton scattering $(\mathrm{p}, \mathrm{p'}\gamma)$}

\author{M.~Elvers}
\email{elvers@ikp.uni-koeln.de}
\ikp
\yale

\author{S.~Pascu}
\email{pascu@ikp.uni-koeln.de}
\ikp
\magu

\author{T.~Ahmed}
\rich

\author{T.~Ahn}
\yale

\author{V.~Anagnostatou}
\surrey
\yale

\author{N.~Cooper}
\yale

\author{C.~Deng}
\rich

\author{J.~Endres}
\ikp

\author{P.~Goddard}
\surrey
\yale

\author{A.~Heinz}
\chalmers
\yale

\author{G.~Ilie}
\yale
\magu

\author{E.~Jiang}
\rich

\author{C.~K\"uppersbusch}
\ikp

\author{D.~Radeck}
\ikp
\yale

\author{D.~Savran}
\emmi
\fias
\yale

\author{N.~Shenkov}
\rich

\author{V.~Werner}
\yale

\author{A.~Zilges}
\ikp

\date{\today}

\begin{abstract}
Octupole vibrational states were studied in the nucleus $^{150}\mathrm{Nd}$ via inelastic proton scattering with $\unit[10.9]{MeV}$ protons which are an excellent probe to excite natural parity states. For the first time in $^{150}\mathrm{Nd}$, both the scattered protons and the $\gamma$ rays were detected in coincidence giving the possibility to measure branching ratios in detail. Using the coincidence technique, the $B(E1)$ ratios of the decaying transitions for 10 octupole vibrational states and other negative-parity states to the yrast band were determined and compared to the Alaga rule. The positive and negative-parity states revealed by this experiment are compared with Interacting Boson Approximation (IBA) calculations performed in the \(spdf\) boson space. The calculations are found to be in good agreement with the experimental data, both for positive and negative-parity states.
\end{abstract}


\maketitle

\section{Introduction}
Octupole vibrations play a fundamental role in the description of low-lying negative-parity states. In a microscopic model, those states are described as a coherent superposition of 1-particle-1-hole excitations between an intruder state and a state of the same major shell with spin difference $\Delta l= \Delta j = 3$ \cite{Butl96}. Collective excitations in vibrational rare earth nuclei, with both proton and neutron numbers in the proximity of a shell closure, are macroscopically described as (multi)phonon excitations \cite{Wilh96}. Further away from shell closures, when prolate quadrupole deformation in the ground state is established, rotational bands are superimposed on the vibrations \cite{Donn66, Neer70a}.\\
The nucleus  $^{150}\mathrm{Nd}$ ranks among the transitional nuclei between spherical harmonic vibrators and axially symmetric rotors \cite{Krue02}. Unlike a vibrator, this nucleus has a spheroidal shape and the quantum number $K$ is rather well defined. Consequently, similar to perfect rotors, a band structure with $\beta$-, $\gamma$-, and an octupole-vibrational band exists in this nucleus \cite{Pitz90}. However, the energy ratio, $E_{4_1^+}/E_{2_1^+}=2.93$, is between a vibrator and a rotor. Furthermore, the relative transition strengths $\frac{B(E2: J^+\rightarrow(J-2)^+)}{B(E2: 2^+\rightarrow0^+)}$ of the yrast band members as a function of the angular momentum $J$ lie between the predictions for an axially symmetric rotor and a vibrator.\\
While the level structure and decay properties of positive-parity states of even spin in the low-energy region are well established in this nucleus, the knowledge concerning low-energy negative-parity states is rather sparse. Up to now, only the decay of the first two members of the $\kp=0^-$ octupole band were observed and the energies of the $\jp=1^-$ to $\jp=7^-$ state were determined by means of inelastic particle scattering $\pp$ \cite{Pign93}. Recently, those states were theoretically described by Bizzeti \textit{et al.}\ using a model valid for nuclear shapes close to axial symmetry \cite{Bizz10}.\\
The goal of this experiment was to determine the ratios of the $B(E1)$ transitions strengths of the remaining two known members in this band and to determine the $K$ quantum number of higher-lying negative-parity states. An approximate agreement with the Alaga rule \cite{Alag55} would further confirm the existence of a spheroidal shape.\\
The structure of \(^{150}\)Nd was interpreted in terms of the first-order phase transition from spherical harmonic vibrator to axially deformed nuclei \cite{Krue02}. The observed agreement revealed that this nucleus is the best case so far of an empirical realization of the X(5) critical point symmetry \cite{Iach01}. However, the X(5) predictions are made only for the yrast and for the \(\beta\)-vibrational bands (or \(s=1\) and \(s=2\) bands in the X(5) description). The model gives no predictions for the \(\gamma\)-vibrational band and, more important for the present discussion, for the octupole-vibrational states. Several attempts have been made in order to describe the negative-parity states for this type of nuclei, for example, the authors of Ref. \cite{Mink06} considered a collective Hamiltonian for the rotation-vibration motion in which the axial quadrupole and octupole degrees of freedom are coupled through the centrifugal interaction, or in Ref. \cite{Bizz10} the collective motion was investigated based on a new parameterization of the octupole and quadrupole degrees of freedom. Another nucleus  proposed to be an empirical realization of the X(5) critical point symmetry (\(^{152}\)Sm) has already been investigated using the IBA-\(spdf\) model approach and a good agreement with the experimental data was concluded \cite{Babi05}. In the present paper, we compare the new experimental results for \(^{150}\)Nd with calculations performed in the framework of the IBA-\(spdf\) model. The reproduction of the \(E1\) transition rates is found to be in good agreement with the predictions given by the IBA calculations, especially for the \(K^{\pi}=0^{-}\) band.\\
The experimental setup and the data analysis are introduced in \secref{sec:setup} and \secref{sec:analysis}, respectively. Thereafter, in \secref{sec:results}, the results are discussed and compared to the theoretical calculations.
\section{Experimental Setup}
\label{sec:setup}
A $\unitfrac[1.5]{mg}{cm^2}$ thick and 98\% enriched self-supporting $\nucl{150}{Nd}$ target was bombarded with a $\unit[10.9]{MeV}$ proton beam provided by the ESTU Tandem Accelerator at the Wright Nuclear Structure Laboratory at Yale University. The experiment lasted for 120 hours at an average beam current of $\unit[4]{nA}$ and an average coincidence rate of about $\unit[2]{kHz}$ $\mathrm{p}\gamma$.\\
For the detection of the scattered protons, five silicon surface-barrier detectors were used covering a solid angle of 5\%. Three of them had a thickness of $\unit[1]{mm}$, two had a thickness of $\unit[5]{mm}$. The elastically scattered protons are completely stopped in both types of detectors. The setup was optimized for high efficiency by placing all detectors as close as possible to the target and it was kept flexible by allowing a positioning of the detectors in all three spacial directions. In order to achieve a high ratio of inelastically to elastically scattered protons, four detectors were placed at $\approx130^\circ$ backward angles where the cross section for elastically scattered protons is orders of magnitudes lower. Using a $\nucl{228}{Th}$ $\alpha$ source, the energy resolution at $\unit[8.784]{MeV}$ was determined ranging from 55 to $\unit[90]{keV}$ for the different detectors while the resolution degraded to values between 110 to $\unit[280]{keV}$ in-beam due to straggling effects in the target and increased electronic noise.\\
The $\gamma$ rays were detected by the YRAST ball spectrometer \cite{Beau00} which is situated at WNSL providing up to 10 segmented HPGe clover detectors. Each detector consists of four crystals with its own preamplified time and energy output. Each clover detector is surrounded by  bismuthe germanate (BGO) and actively shielded. The total photopeak efficiency of the array is in the order of $2\%$ at $\unit[1332]{keV}$ with an average energy resolution of \unit[2]{keV}. In the present experiment in total nine detectors were used while two were placed at $45^\circ$ with respect to the beam axis, five at $90^\circ$ and two at $135^\circ$. The beam was stopped by a carbon and lead shielded beam dump $\unit[2]{m}$ downstream which strongly reduced the neutron and $\gamma$-ray background. The data acquisition was done by three 16-channel ADCs, three 16-channel TDCs and a logic unit based on VME electronics. Data was acquired using three triggers: $p\gamma$ coincidences, downscaled single HPGe and downscaled silicon events.
\section{Analysis}
\label{sec:analysis}
\begin{figure*}
\includegraphics[width=\textwidth]{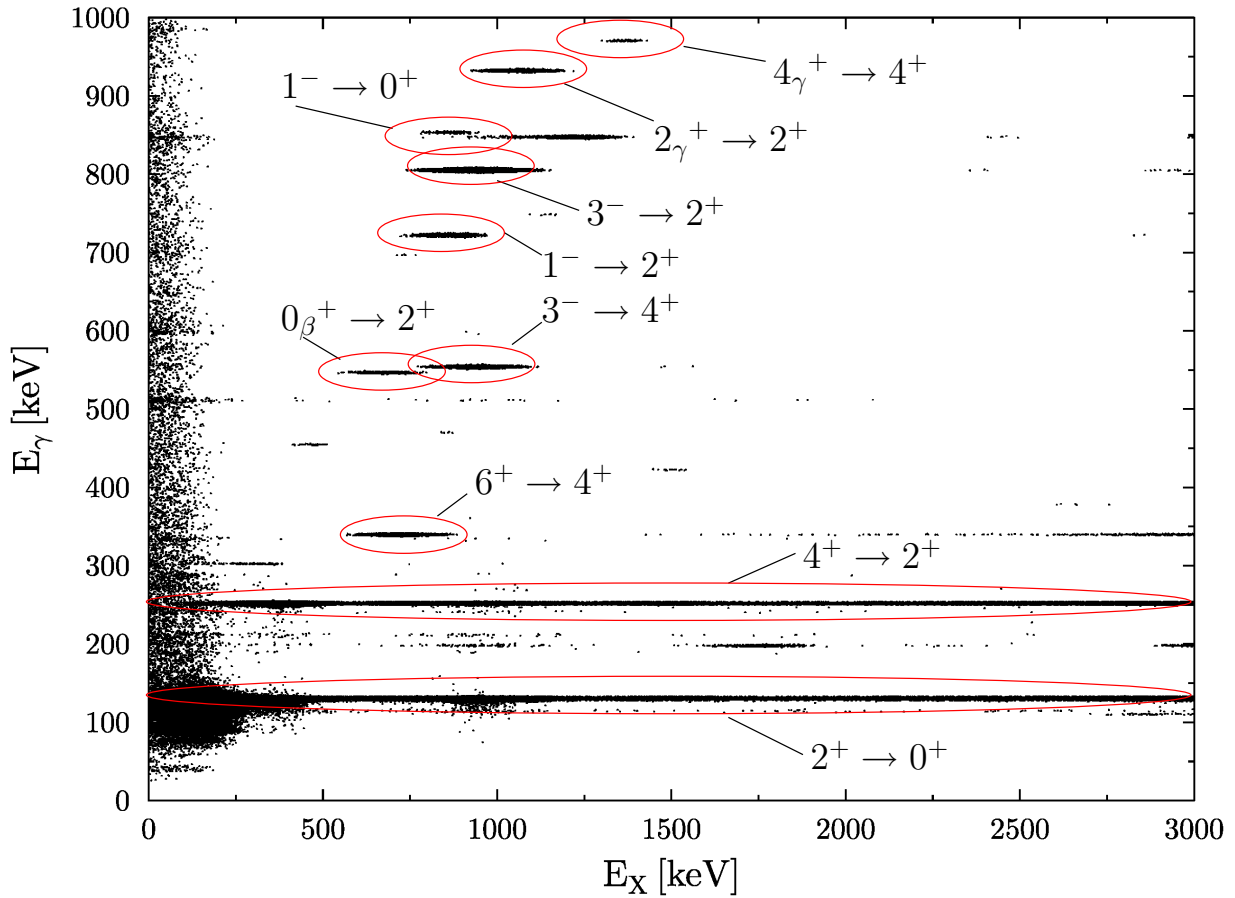}
\caption{\label{fig:matrix} The $\mathrm{p}\gamma$ matrix of $\nucl{150}{Nd}$. The transition energy $E_\gamma$ is plotted versus the excitation energy $E_X$. Due to the worse energy resolution in the silicon detectors in comparison to the HPGe detectors, transitions appear as narrow lines. This graphical representation of the $\mathrm{p}\gamma$ matrix nicely shows the strength of the method to investigate the decay pattern of excited states. }
\end{figure*}
\begin{figure*}
\includegraphics[width=\textwidth]{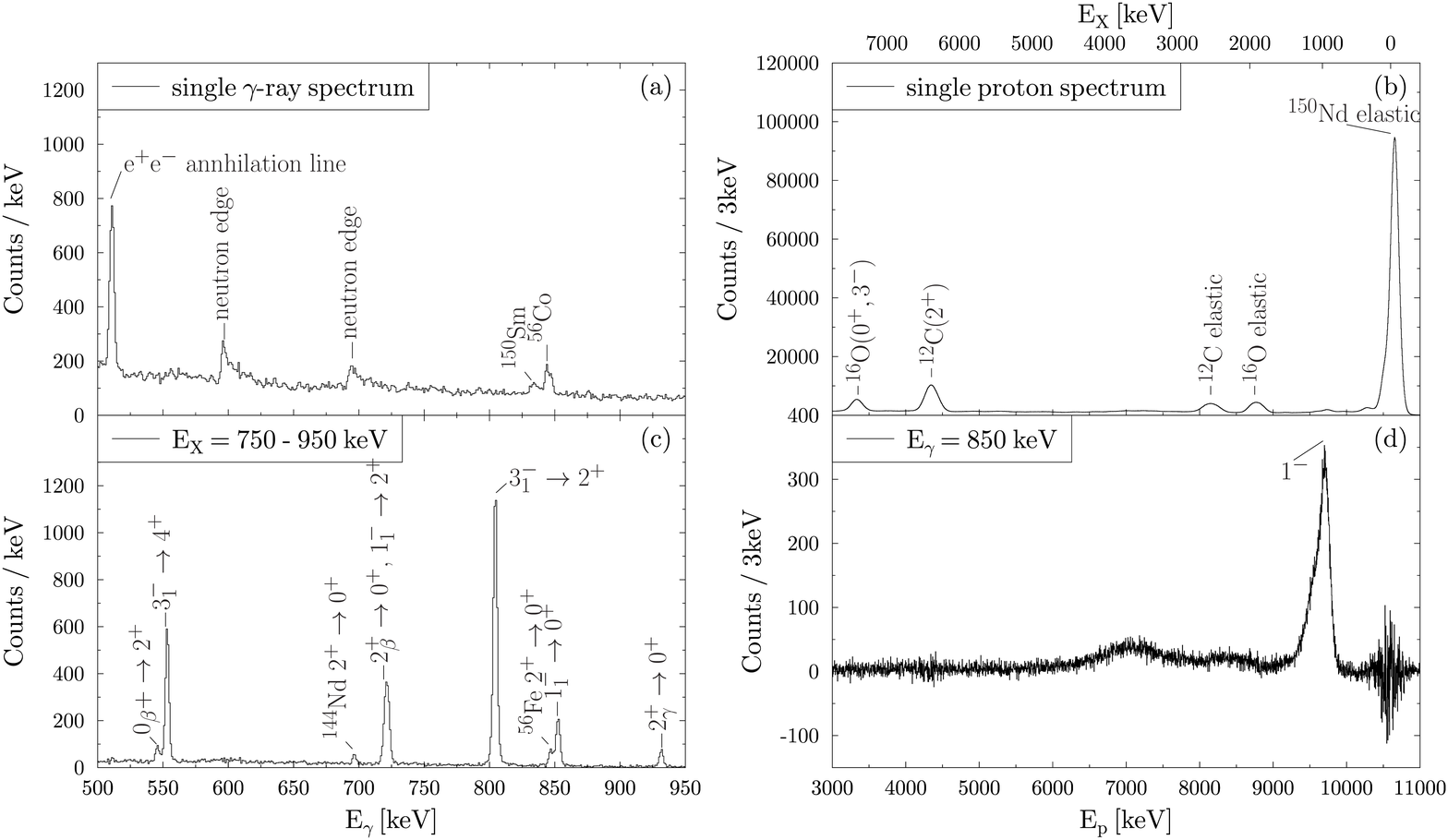}
\caption{\label{fig:cuts} Single and coincidence spectra of a silicon detector and a clover detector. The coincidence spectra were corrected for random coincidences. The single spectra are dominated by background reactions while the gated spectra clearly show the transitions within $\nucl{150}{Nd}$. A clover single spectrum (one leaf, $\unit[1]{h}$ run and a single silicon spectrum (all runs) are shown in panels (a) and (b), respectively, while the gated spectra (all runs) are shown in panels (c) and (d).}
\end{figure*}

The single leaves of each clover were calibrated in energy and their relative efficiencies were determined using a $\nucl{152}{Eu}$ source placed at target position. The silicon detectors were calibrated using peaks in the proton single spectra corresponding to elastic scattering and to excitations of the first excited states of $\nucl{12}{C}$, $\nucl{16}{O}$ and $\nucl{150}{Nd}$ measured in the in-beam spectra. In a next step, the energy shift factors were determined and the prompt and random windows were applied to the time difference spectra in order to reduce the background. In order to further reduce the Compton background in the lower energy part and to increase the photo-peak efficiency, the energies in the four leaves of a clover detector were added eventwise using an addback algorithm \cite{Josh97}.\\
Using the sorting code \textit{CSCAN} \cite{Cscan04}, the proton-$\gamma$-coincidence matrix ($\mathrm{p}\gamma$ matrix), all single spectra, and the $\gamma$-$\gamma$-coincidence matrix ($\gamma\gamma$ matrix) were generated. An excerpt of the $\mathrm{p}\gamma$ matrix is shown in \figref{fig:matrix}. All transitions appear as narrow lines due to the worse energy resolution of the silicon detectors in comparison to the good energy resolution of HPGe detectors. Both the $2^+\rightarrow0^+$ and the $4^+\rightarrow2^+$ transition cover the entire range of the excitation energies $E_X$ giving rise to the fact that most of the states decay via those states to the groundstate.\\
The necessity to detect both the scattered proton and the $\gamma$ transition is illustrated in \figref{fig:cuts}. Without coincidence condition, the $\gamma$-ray spectrum (a) shows the $e^+e^-$ annihilation line and the neutron edges as main contributions. In particular, no indication for transitions of states in $\nucl{150}{Nd}$ are visible. However, when a gate with $\unit[750]{keV}\le E_X \le \unit[950]{keV}$ is applied in the proton spectrum, peaks corresponding to the decay of states in $\nucl{150}{Nd}$ located at this excitation energy are clearly visible (c). Similarly, the particle spectrum is dominated by the elastic lines of $\nucl{12}{C}$, $\nucl{16}{O}$ and $\nucl{150}{Nd}$ and their first excited states (b). If a gate is applied at a transition energy of the first $\jp=1^-$ state in the $\gamma$-ray spectrum, a peak corresponding to the energy of the $\jp=1^-$ state is clearly visible in the silicon spectrum (d). The fluctuations at $\unit[10.5]{MeV}$ are the remnants of the elastic line of $\nucl{150}{Nd}$.\\

\section{Results and Discussion}
\label{sec:results}
Using the $\mathrm{p}\gamma$-coincidence matrix, the level scheme of $\nucl{150}{Nd}$ was corrected and extended. In total, 37 levels up to $\unit[2]{MeV}$ were excited and 64 transitions were observed. In particular, the transitions of 8 states being previously excited in inelastic particle scattering experiments were found \cite{Pign93} and 12 new states were observed for the first time. Table \ref{tab:states} summarizes the results on $\nucl{150}{Nd}$ obtained from the present experiment.\\

\begin{longtable}{r@{.} l l l l l l r@{.} l r@{.} l r@{(} l }
\caption{States up to $\unit[2]{MeV}$ and their decay observed in the $\nucl{150}{Nd}\ppg$ experiment. Octupole states strongly decaying to the $\beta$-vibrational band are marked with $\mathcal{O}_\beta$. The state energy $E$, the initial and final spin and K quantum number $J_{i,f}^\pi$  $K_{i,f}^\pi$, the final energy $E_f$, the energy of the $\gamma$-ray $E_\gamma$ and the branching ratios $I_\gamma$ are listed.}\\
\hline\hline
\multicolumn{2}{l}{$E$ [keV]} & $K_i^\pi$ & $J_i^\pi$ &  & $J_f^\pi$ & $K_f^\pi$ & \multicolumn{2}{l}{$E_f$ [keV]} & \multicolumn{2}{l}{$E_\gamma$ [keV]} & \multicolumn{2}{l}{$I_\gamma$}\\
\hline
\endfirsthead
\hline\hline

\multicolumn{2}{l}{$E$ [keV]} & $K_i^\pi$ & $J_i^\pi$ &  & $J_f^\pi$ & $K_f^\pi$ & \multicolumn{2}{l}{$E_f$ [keV]} & \multicolumn{2}{l}{$E_\gamma$ [keV]} & \multicolumn{2}{l}{$I_\gamma$}\\
\hline
\endhead
\hline
\multicolumn{13}{l}{{Continued on next page}} \\
\endfoot
\hline
\multicolumn{13}{l}{$^a$ observed for the first time, not in NNDC} \\
\multicolumn{13}{l}{$^b$ state without transitions observed in inelastic particle scattering} \\
\multicolumn{13}{l}{$^c$ Not excited, taken from NNDC} \\
\endlastfoot
130&21(8) & $0_{1}^+$&$2^+$ & $\rightarrow$ & $0^+$& & 0&00(0) & 130&2 & \multicolumn{2}{l}{100.0}\\
\label{tab:states}
381&45(11) & $0_{1}^+$&$4^+$ & $\rightarrow$ & $2^+$&$0_{1}^+$ & 130&21(8) & 251&24 & \multicolumn{2}{l}{100.0}\\
675&37(17) & $0_\beta^+$&$0^+$ & $\rightarrow$ & $2^+$&$0_{1}^+$ & 130&21(8) & 546&06 & \multicolumn{2}{l}{100.0}\\
720&40(50) & $0_{1}^+$&$6^+$ & $\rightarrow$ & $4^+$&$0_{1}^+$ & 381&45(11) & 339&70 & \multicolumn{2}{l}{100.0}\\
850&79(21) & $0_\beta^+$&$2^+$ & $\rightarrow$ & $4^+$&$0_{1}^+$ & 381&45(11) & 469&40 & 28.8&40)\\\nopagebreak
\multicolumn{2}{l}{} & & & & $2^+$&$0_{1}^+$ & 130&21(8) & 720&61 & 100.0&130)\\\nopagebreak
\multicolumn{2}{l}{} & & & & $0^+$& & 0&00(0) & 850&70 & 14.5&64)\\
853&06(21) & $0^-$&$1^-$ & $\rightarrow$ & $2^+$&$0_{1}^+$ & 130&21(8) & 722&91 & 100.0&134)\\\nopagebreak
\multicolumn{2}{l}{} & & & & $0^+$& & 0&00(0) & 853&01 & 97.7&165)\\
934&96(20) & $0^-$&$3^-$ & $\rightarrow$ & $4^+$&$0_{1}^+$ & 381&45(11) & 553&52 & 36.3&39)\\\nopagebreak
\multicolumn{2}{l}{} & & & & $2^+$&$0_{1}^+$ & 130&21(8) & 804&73 & 100.0&104)\\
1061&99(20) & $2_\gamma^+$&$2^+$ & $\rightarrow$ & $4^+$&$0_{1}^+$ & 381&45(11) & 680&29 & 5.5&8)\\\nopagebreak
\multicolumn{2}{l}{} & & & & $2^+$&$0_{1}^+$ & 130&21(8) & 931&78 & 100.0&106)\\\nopagebreak
\multicolumn{2}{l}{} & & & & $0^+$& & 0&00(0) & 1061&98 & 80.4&88)\\
1129&57(21) & $0^-$&$5^-$ & $\rightarrow$ & $6^+$&$0_{1}^+$ & 720&40(50) & 409&23$^a$ & 24.3&37)\\\nopagebreak
\multicolumn{2}{l}{} & & & & $4^+$&$0_{1}^+$ & 381&45(11) & 748&07$^a$ & 100.0&118)\\
1129&70(110) & $0_{1}^+$&$8^+$ & $\rightarrow$ & $6^+$&$0_{1}^+$ & 720&40(50) & 409&50$^c$ & \multicolumn{2}{l}{100.0}\\
1137&59(41) & $0_\beta^+$&$4^+$ & $\rightarrow$ & $6^+$&$0_{1}^+$ & 720&40(50) & 416&94$^a$ & 43.5&76)\\\nopagebreak
\multicolumn{2}{l}{} & & & & $4^+$&$0_{1}^+$ & 381&45(11) & 756&40$^a$ & 100.0&143)\\
1183&33(21) & $ $&$\le2$ & $\rightarrow$ & $3^-$&$0^-$ & 934&96(20) & 248&34$^a$ & 100.0&142)\\\nopagebreak
\multicolumn{2}{l}{} & & & & $1^-$&$0^-$ & 853&06(21) & 330&30 & 66.1&102)\\
1200&49(27) & $ $&$3(^-)$ & $\rightarrow$ & $4^+$&$0_{1}^+$ & 381&45(11) & 819&16 & 22.4&37)\\\nopagebreak
\multicolumn{2}{l}{} & & & & $2^+$&$0_{1}^+$ & 130&21(8) & 1070&15 & 100.0&118)\\
1284&59(20) & $ $&$1^-$ & $\rightarrow$ & $3^-$&$0^-$ & 934&96(20) & 349&63 & 67.8&103)\\\nopagebreak
\multicolumn{2}{l}{} & & & & $1^-$&$0^-$ & 853&06(21) & 430&88 & 100.0&132)\\\nopagebreak
\multicolumn{2}{l}{} & & & & $2^+$&$0_\beta^+$ & 850&79(21) & 434&09 & 9.8&36)\\
1351&57(21)$^b$ & $2_\gamma^+$&$4^+$ & $\rightarrow$ & $2^+$&$2_\gamma^+$ & 1061&99(20) & 289&57$^a$ & 6.4&10)\\\nopagebreak
\multicolumn{2}{l}{} & & & & $4^+$&$0_{1}^+$ & 381&45(11) & 970&07$^a$ & 100.0&112)\\\nopagebreak
\multicolumn{2}{l}{} & & & & $2^+$&$0_{1}^+$ & 130&21(8) & 1221&42$^a$ & 38.7&49)\\
1432&94(24) & $0^-$&$(7^-)$ & $\rightarrow$ & $8^+$&$0_{1}^+$ & 1129&70(110) & 303&33$^a$ & 49.6&137)\\\nopagebreak
\multicolumn{2}{l}{} & & & & $6^+$&$0_{1}^+$ & 720&40(50) & 712&44 & 100.0&223)\\
1435&90(20) & $$&$4^+$ & $\rightarrow$ & $3(^-)$&$$ & 1200&49(27) & 234&41 & \multicolumn{2}{l}{100.0}\\\nopagebreak
\multicolumn{2}{l}{} & & & & $2^+$&$2_\gamma^+$ & 1061&99(20) & 373&91 & \multicolumn{2}{l}{100.0}\\
1483&67(21)$^b$ & $$&$3^-$ & $\rightarrow$ & $3(^-)$&$$ & 1200&49(27) & 283&00$^a$ & 31.7&44)\\\nopagebreak
\multicolumn{2}{l}{} & & & & $2^+$&$2_\gamma^+$ & 1061&99(20) & 421&71$^a$ & 33.7&44)\\\nopagebreak
\multicolumn{2}{l}{} & & & & $4^+$&$0_{1}^+$ & 381&45(11) & 1102&14$^a$ & 100.0&120)\\\nopagebreak
\multicolumn{2}{l}{} & & & & $2^+$&$0_{1}^+$ & 130&21(8) & 1353&52$^a$ & 85.9&113)\\
1488&52(20)$^a$ & $$&$4, (0, 2)$ & $\rightarrow$ & $1^-$&$$ & 1284&59(20) & 204&15$^a$ & 31.4&74)\\\nopagebreak
\multicolumn{2}{l}{} & & & & $2^+$&$0_{1}^+$ & 130&21(8) & 1358&31$^a$ & 100.0&159)\\
1489&29(20)$^a$ & $ $&$0,1,2$ & $\rightarrow$ & $0^+$&$0_\beta^+$ & 675&37(17) & 813&92$^a$ & \multicolumn{2}{l}{100.0}\\
1496&75(20)$^a$ & $ $&$3, (2,4,5)$ & $\rightarrow$ & $4^+$&$0_\beta^+$ & 1137&59(41) & 359&16$^a$ & \multicolumn{2}{l}{100.0}\\
1517&67(22)$^a$ & $ $&$4,5,6$ & $\rightarrow$ & $6^+$&$0_{1}^+$ & 720&40(50) & 797&21$^a$ & 69.4&145)\\\nopagebreak
\multicolumn{2}{l}{} & & & & $4^+$&$0_{1}^+$ & 381&45(11) & 1136&29$^a$ & 100.0&160)\\
1545&06(36)$^b$ & $\mathcal{O}_\beta$&$3^-$ & $\rightarrow$ & $2^+$&$0_\beta^+$ & 850&79(21) & 694&05$^a$ & 25.0&71)\\\nopagebreak
\multicolumn{2}{l}{} & & & & $2^+$&$0_{1}^+$ & 130&21(8) & 1415&06$^a$ & 100.0&149)\\
1580&10(30)$^b$ & $ $&$3^-$ & $\rightarrow$ & $4^+$&$0_{1}^+$ & 381&45(11) & 1198&50$^a$ & 100.0&121)\\\nopagebreak
\multicolumn{2}{l}{} & & & & $2^+$&$0_{1}^+$ & 130&21(8) & 1450&05$^a$ & 67.5&93)\\
1645&34(26)$^a$ & $ $&$5, (4)$ & $\rightarrow$ & $6^+$&$0_{1}^+$ & 720&40(50) & 924&82$^a$ & 57.0&93)\\\nopagebreak
\multicolumn{2}{l}{} & & & & $4^+$&$0_{1}^+$ & 381&45(11) & 1264&01$^a$ & 100.0&138)\\
1645&97(26)$^a$ & $ $&$3, 5, (2,4)$ & $\rightarrow$ & $4^+$&$2_\gamma^+$ & 1351&57(21) & 294&29$^a$ & 100.0&159)\\\nopagebreak
\multicolumn{2}{l}{} & & & & $4^+$&$0_\beta^+$ & 1137&59(41) & 508&50$^a$ & 34.5&88)\\
1648&66(20)$^a$ & $ $&$1, (0, 2)$ & $\rightarrow$ & $2^+$&$0_{1}^+$ & 130&21(8) & 1518&45$^a$ & \multicolumn{2}{l}{100.0}\\
1738&86(20) & $ $&$0^+$ & $\rightarrow$ & $2^+$&$0_{1}^+$ & 130&21(8) & 1608&65 & \multicolumn{2}{l}{100.0}\\
1765&28(22)$^a$ & $ $&$0,1,2,3$ & $\rightarrow$ & $1^-$&$0^-$ & 853&06(21) & 912&16$^a$ & 100.0&140)\\\nopagebreak
\multicolumn{2}{l}{} & & & & $1^-$&$ $ & 1284&59(20) & 480&75$^a$ & 42.1&90)\\
1777&08(20)$^a$ & $$&$1,2,3,4,5$ & $\rightarrow$ & $3^-$&$$ & 1580&10(30) & 196&98$^a$ & 100.0&0)\\
1782&21(20)$^b$ & $ $&$(4^+)$ & $\rightarrow$ & $3^-$&$0^-$ & 934&96(20) & 847&25$^a$ & 100.0&0)\\
1799&91(20)$^b$ & $ $&$(5^-)$ & $\rightarrow$ & $3^-$&$0^-$ & 934&96(20) & 864&22$^a$ & 100.0&190)\\\nopagebreak
\multicolumn{2}{l}{} & & & & $6^+$&$0_{1}^+$ & 720&40(50) & 1080&23$^a$ & 15.2&74)\\
1864&20(20)$^b$ & $\mathcal{O}_\beta$&$3^-$ & $\rightarrow$ & $2^+$&$0_\beta^+$ & 850&79(21) & 1014&08$^a$ & 100.0&126)\\\nopagebreak
\multicolumn{2}{l}{} & & & & $4^+$&$0_{1}^+$ & 381&45(11) & 1482&75$^a$ & 80.7&112)\\
1906&52(20)$^b$ & $ $&$4^+$ & $\rightarrow$ & $2^+$&$2_\gamma^+$ & 1061&99(20) & 844&53$^a$ & 100.0&0)\\
1909&34(20)$^a$ & $ $&$0,1,2,3,4$ & $\rightarrow$ & $2^+$&$0_{1}^+$ & 130&21(8) & 1779&13$^a$ & 100.0&0)\\
1975&62(20)$^a$ & $ $&$1,2,3,4,5$ & $\rightarrow$ & $3(^-)$&$ $ & 1200&49(27) & 775&13$^a$ & 100.0&0)\\
1984&94(20)$^a$ & $ $&$1,2,3,4,5$ & $\rightarrow$ & $3^-$&$0^-$ & 934&96(20) & 1049&98$^a$ & 100.0&0)\\
\end{longtable}

In the $\kp=0^-$ band, the first four members ($1^-$, $3^-$, $5^-$, and $7^-$) were excited and branching ratios of the latter two were determined for the first time. The experimental branching ratios and the predictions by the Alaga rule are listed in \tabref{Table1} showing a good agreement for the first three states and an underprediction for the $\jp=7^-$ state by the Alaga rule. The approximate agreement with the Alaga rule for the first three band members is another strong indication for an axially deformed rotor with a separable rotational motion.\\
The transitions of higher-lying negative-parity states were observed for the first time. Similar to the states described above, most of the states have $\kp=0^-$ character according to the Alaga rule which is shown in \tabref{Table1}. Two states show a significant branching to the $0_\beta^+$ vibrational band giving rise to the assumption that these states have a large component of the $\beta$-vibrational in their wave function. The energies of those two candidates are located at $\unit[1545]{keV}$ and $\unit[1864]{keV}$ which is in the region of the sum energy $E_\mathrm{sum}=E_{\mathrm{oct},1^-}+E_{\beta, 2^+}=\unit[1704]{keV}$ as predicted in \cite{Donn66}. All corresponding states are marked with $\mathcal{O}_\beta$ in \tabref{tab:states}.\\
The interacting boson model \cite{Iac75} provides a phenomenological approach for studying nuclear structure by introducing bosons of a given spin. Each type of bosons is associated with a corresponding multipole mode. The quadrupole vibration and deformation are described in terms of \(s\) and \(d\) bosons (\(L=0\) and \(L=2\)), while negative parity states are described by introducing \(p\) and \(f\) bosons (\(L=1\) and \(L=3\)). The \(f\) boson has been introduced in order to describe the octupole vibrational bands in quadrupole deformed nuclei \cite{Barf88,Cott96,Cott98}. The physical nature of the \(p\) boson is not yet very clear. In Ref. \cite{Enge85} it was assumed that the origin of the \(p\) boson could be a possible correction of the spurious center of mass motion of the quadrupole-octupole intrinsic system and in Ref. \cite{Sugi96} the \(p\) boson was related to the giant dipole resonance. Despite these studies, the physical meaning of the \(p\) boson remains an open question.

 Calculations were performed in the \(spdf\) IBA-1 framework (no distinction was made between protons and neutrons) using the Extended Consistent Q-formalism (ECQF) \cite{Cast88}. The Hamiltonian employed  in the present paper is the natural extension of the \(\hat{H}_{sd}\) Hamiltonian and is able to describe simultaneously the positive and negative parity states:

\begin{eqnarray}
\hat{H}_{spdf}=\mathrm{\epsilon}_{d} \hat{n}_{d}+\mathrm{\epsilon}_{p} \hat{n}_{p}+\mathrm{\epsilon}_{f} \hat{n}_{f}+\mathrm{\kappa}(\hat{Q}_{spdf}\cdot \hat{Q}_{spdf})^{(0)}\nonumber\\
+ \mathrm{a_{3}}[(\hat{d}^{\dagger}\tilde{d})^{(3)} \cdot (\hat{d}^{\dagger}\tilde{d})^{(3)}]^{(0)}+\mathrm{a_{4}}[(\hat{d}^{\dagger}\tilde{d})^{(4)} \cdot (\hat{d}^{\dagger}\tilde{d})^{(4)}]^{(0)}\label{eq1}
\end{eqnarray}
where \(\epsilon_{d}\), \(\epsilon_{p}\), and \(\epsilon_{f}\) are the boson energies and \(\hat{n}_{p}\), \(\hat{n}_{d}\) and \(\hat{n}_{f}\) are the boson number operators. In the \(spdf\) model, the quadrupole operator is considered as being \cite{Kusn90}:
\begin{eqnarray}
\hat{Q}_{spdf}=\hat{Q}_{sd}+\hat{Q}_{pf}=[{(\hat{s}^{\dagger}\tilde{d}+\hat{d}^{\dagger}\hat{s})^{(2)}-\frac{\sqrt{7}}{2} (\hat{d}^{\dagger}\tilde{d})^{(2)}}]\nonumber\\
-\frac{3\sqrt{7}}{5}[(p^{\dagger}\tilde{f}+f^{\dagger}\tilde{p})]^{(2)}+\frac{9\sqrt{3}}{10}(p^{\dagger}\tilde{p})^{(2)}+\frac{3\sqrt{42}}{10}(f^{\dagger}\tilde{f})^{(2)}.\label{eq2}
\end{eqnarray}

The quadrupole electromagnetic transition operator is:
\begin{eqnarray}
\hat{T}(E2)=e_{2} \hat{Q}_{spdf}\label{eq3}
\end{eqnarray}
where \(e_{2}\) represents the boson effective charge.

This form of the quadrupole operator implies to use the same quadrupole parameter strength \(\kappa\) to describe \(sd\) bosons and \(pf\) bosons. In addition, the rotational SU\(_{spdf}\)(3) structure is also generated by this form [Eq.\,(\ref{eq2})] of the \(\hat{Q}_{spdf}\) quadrupole operator \cite{Kusn90}. In this way no additional free parameters are used in the calculations. The structure of both the positive and negative-parity states is also described by the same parameter-free quadrupole operator \(\hat{Q}_{spdf}\). The only degree of freedom in describing the negative parity states lies in the \(p\) and \(f\) vibrational energies \(\epsilon_{p}\) and \(\epsilon_{f}\). However, these are defined largely by the experimental locations of the lowest 1\(^{-}\) and 3\(^{-}\) states.

For the \(E1\) transitions there is more than one operator in the \(spdf\) algebra. Consequently, a linear combination of the three allowed one-body interactions was taken:
\begin{eqnarray}
\hat{T}(E1)=e_{1}[\chi_{sp}^{(1)}({s}^{\dagger}\tilde{p}+{p}^{\dagger}\tilde{s})^{(1)}+({p}^{\dagger}\tilde{d}+{d}^{\dagger}\tilde{p})^{(1)}\nonumber\\
+\chi_{df}^{(1)}({d}^{\dagger}\tilde{f}+{f}^{\dagger}\tilde{d})^{(1)}]\label{eq4}
\end{eqnarray}
where \(e_{1}\) is the effective charge for the \(E1\) transitions and \(\chi_{sp}^{(1)}\) and \(\chi_{df}^{(1)}\) are two model parameters.

The calculations were performed using the computer code OCTUPOLE \cite{Kusn00}. The Hamiltonian is diagonalized in a Hilbert space with a total number of bosons \(N_{B}=n_{s}+n_{d}+n_{p}+n_{f}\). For the present calculations we used an extended basis allowing up to three negative parity bosons (\(n_{p}+n_{f}\)=3). The vibrational strengths used in the calculations are \(\epsilon_{d}\)=0.53 MeV, \(\epsilon_{p}\)=1.32 MeV, and \(\epsilon_{f}\)=1.12 MeV while the quadrupole-quadrupole interaction strength has a value of \(\kappa\)=-18.5 keV. The other two terms introduced in the Hamiltonian (the octupole and the hexadecapole terms) are included only to obtain a better reproduction for the relative inter- and intraband transitional rates. The strengths of these two terms are set to values of \(a_{3}\)=0.04 MeV and \(a_{4}\)=0.10 MeV, respectively.

\begin{figure*}
\centering
\includegraphics[width=13.5cm]{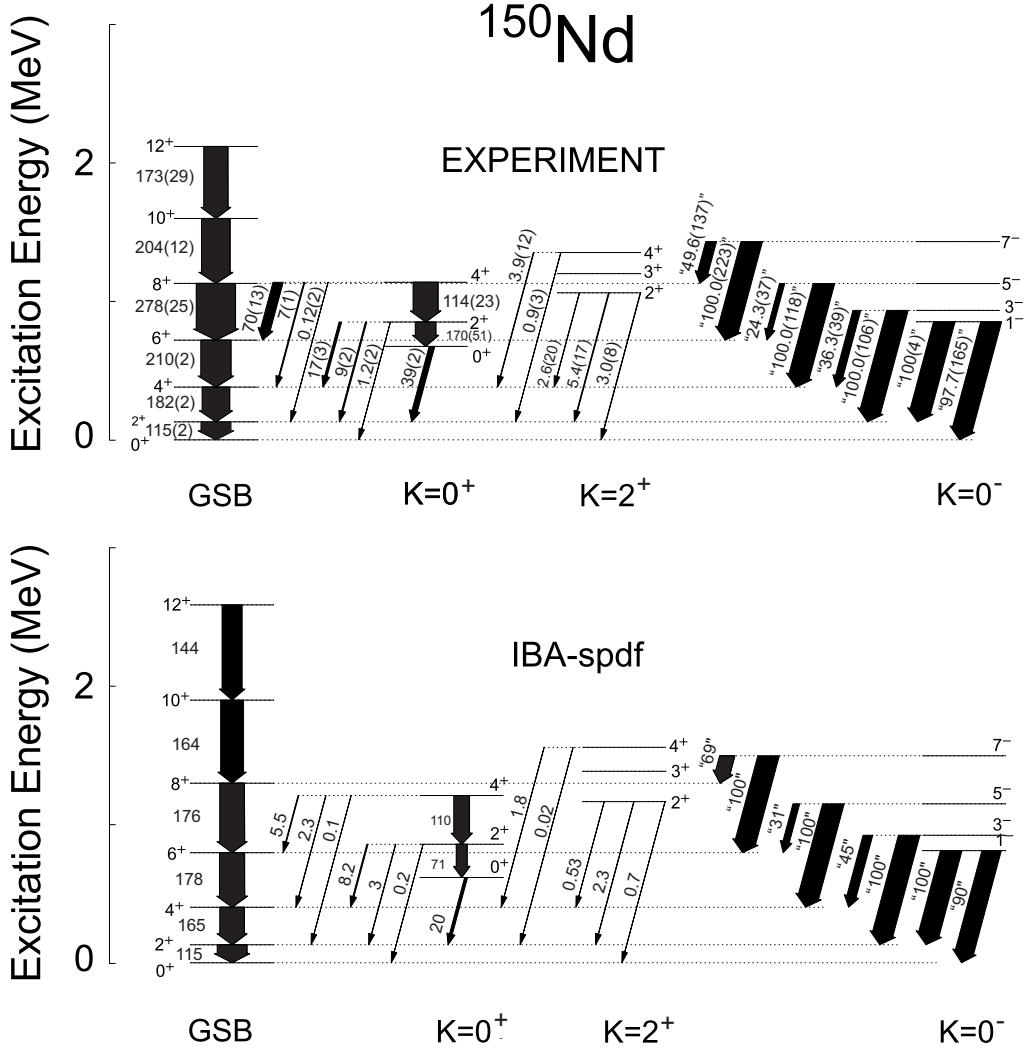}
\caption{Comparison between experimental (top) and calculated (bottom) excitation energies and transition probabilities or relative intensities of the \(\gamma\)-ray transitions for the low-lying levels in \(^{150}\)Nd. The known \(B(E2)\) values are indicated for the positive-parity states. The numbers in quotation marks are the relative values (branching ratios) and are indicated for the negative-parity states. The experimental information is taken from Refs. \cite{Krue02,Ziel04,Derm95} and from the present work. The bands are labeled according to their \(K\) quantum number.}
\label{fig1}
\end{figure*}

The quality of the present calculation in describing the experimental data can be seen in Fig.\ref{fig1}. The numbers on the transition arrows represent the absolute \(B(E2)\) values given in Weisskopf Units (W.u.) and are indicated for the positive parity states, and relative intensities of the \(\gamma\)-ray transitions (branching ratios) given for the negative parity states (only the lifetime of the 1\(^{-}\) state is measured \cite{Ziel04}, hence only one absolute \(B(E1)\) value is known). The overall agreement between the experimental and calculated energies is good. The spacing of the states is reasonably well reproduced by the calculations for all the bands involved in our analysis (the ground state band, the \(\beta\)- and \(\gamma\)-vibrational bands and the \(K^{\pi}=0^{-}\) octupole vibrational band). The first excited 0\(^{+}\) state is located less than 50 keV below the experimental one with an energy of 675.4 keV \cite{Derm95}. The location of this state is considered a key prediction in the X(5) model because the ratio \(E(0^{+}_{2})/E(2^{+}_{1})\) is fixed at a value of 5.67 \cite{Cast06} (cannot be fitted to obtain a better agreement) and gives an important fingerprint for the presence of the first-order critical point symmetry. The relative values of the \(E1\) and \(E2\) transition probabilities  are in very good agreement with the experiment as can be seen in Fig.\ref{fig1} for the negative parity states (only \(E1\) transitions). However, the absolute \(B(E2)\) values are in general underestimated by a factor that usually does not exceed a value of two. Especially for the higher spin states, the finite boson number effect plays an important role in lowering the predictions for \(E2\) transition strengths. This behavior was already observed in Ref. \cite{Mccu05}, where a detailed comparison between X(5) critical point symmetry and the IBA model predictions for this region was made.

For the \(E2\) transitions we obtained the agreement from Fig.\ref{fig1} by using a value of \(e_{2}=0.13 \,eb\) for the effective charge. For the \(E1\) transitions values of \(\chi_{sp}\)=-1.31 and \(\chi_{df}\)=0.77 were used in the fit to account for the observed \(E1\) transition rates. The effective charge is not taken into consideration in the present calculations because only the lifetime of the first 1\(^{-}\) state is measured and the extracted \(E1\) absolute strength can only be used for normalization. For the moment no other absolute \(E1\) strengths are experimentally known for the negative parity states and no comparison with the calculated values can be made. Our discussion is based only on the relative values obtained in the present experiment. A comparison between the experimental \(B(E1)\) ratios with the calculated ones is given in Table \ref{Table1}. In addition, the predictions of the Alaga rule are given for \(K^{\pi}\)=\(0^{-}\) and for \(K^{\pi}\)=\(1^{-}\) bands. For the \(K^{\pi}\)=\(0^{-}\) band the agreement of the IBA calculations with the experiment is remarkably good, both for the energies and for the relative transition probabilities. Most of the IBA predictions are within the experimental uncertainties or very close to these values. The predictions of the Alaga rule are in good agreement with the experiment for the first three states (\(J^{\pi}\)=\(1^{-}\), \(3^{-}\), and \(5^{-}\)) and show an underprediction for the last level (\(J^{\pi}\)=\(7^{-}\)) of the \(K^{\pi}=0^{-}\) band. For the states belonging to higher \(K^{\pi}\) some additional assumptions about the spin-parity of the states had to be made. Namely, we assumed that the levels at 1517.7 keV and 1645.3 keV have a \(J^{\pi}\)=\(5^{-}\) spin assignment. Also, a state located at 1200 keV was identified as the \(J^{\pi}\)=\(3^{+}\) member of the \(\gamma\)-vibrational band in an (n,n'\(\gamma\)) experiment \cite{Trip77} while it was assigned as a \(J^{\pi}\)=\(3^{\pm}\) doublet within the scope of a (d,d'\(\gamma\)) experiment \cite{Kata80}. Considering this ambiguity, this level was considered as having a positive parity in the present discussion, although only the energy of this level is given in the comparison presented in Fig.\ref{fig1}. For the other states in Table \ref{Table1} the experimental data are very well described by the IBA calculations, while underestimated, in general, by the Alaga rule predictions. Additional calculations are needed in order to gain a deeper understanding from a microscopic point of view of the electric dipole transitions.

\begin{table*}
\caption{\label{Table1} \(B(E1\)) ratios \(R_{exp}=\frac{B(E1;J^{-}_{i}\to(J_{i}+1)^{\pi})}{B(E1;J^{-}_{i}\to(J_{i}-1)^{\pi})}\) to the yrast band of octupole-vibrational states and negative parity states compared with IBA-\(spdf\) calculations and with the Alaga-rule predictions. Furthermore, the experimental energy $E_{exp}$, the initial spin $J^{\pi}_{i}$, the final $K$ quantum number $K^{\pi}_{f}$ and the IBA calculations for the energy and branching ratio $E_{IBA-spdf} [keV]$ and $R_{IBA-spdf}$ are given. Most of the states are in good agreement with the IBA calculations but differ considerably from the Alaga rule predictions. Members of the \(K^{\pi}=0^{-}\) band are marked in bold font.}
\begin{ruledtabular}
\begin{tabular}{cccccccc}
\vspace{0.1cm}
&&&&&&\multicolumn{2}{c}{Alaga rule}\\
\vspace{0.1cm}
$E_{exp}$ [keV] & $J^{\pi}_{i}$ & $K^{\pi}_{f}$ & $R_{exp}$ & $E_{IBA-spdf} [keV]$ & $R_{IBA-spdf}$ & $K^{\pi}=0^{-}$ & $K^{\pi}=1^{-}$\\ 
\hline
\vspace{0.1cm}
{\bf 853.1}   &  $1^{-}$  &  $0^{+}_{1}$  &  1.68(36)   &  810.1    &  1.82  &  2.00  &  0.50  \\
\vspace{0.1cm}
{\bf 935.0}   &  $3^{-}$  &  $0^{+}_{1}$  &  1.12(17)   &  921.2    &  1.38  &  1.33  &  0.75  \\
\vspace{0.1cm}
{\bf 1129.6} &  $5^{-}$  &  $0^{+}_{1}$  &  1.48(29)   &  1148.0  &  1.89  &  1.20  &  0.83  \\
\vspace{0.1cm}
{\bf 1432.9} &  $7^{-}$  &  $0^{+}_{1}$  &  6.43(228)  &  1494.8  &  4.2   &  1.14  &  0.88  \\
\vspace{0.1cm}
1483.7 &  $3^{-}$  &  $0^{+}_{1}$  &  2.16(38)    &  1414.7  &  1.77  &  1.33  &  0.75  \\
\vspace{0.1cm}
1517.7 &  4,5,6   &  $0^{+}_{1}$  &  2.01(53)        &  1517.7  &  1.1    &  1.20  &  0.83  \\
\vspace{0.1cm}
1580.1 &  $3^{-}$  &  $0^{+}_{1}$  &  2.62(48)    &  1694.5  &  2.17  &  1.33  &  0.75  \\
\vspace{0.1cm}
1645.3 &  4,5,6  &  $0^{+}_{1}$  &  1.46(31)         &  1986.8  &  1.15  &  1.20  &  0.83  \\
\end{tabular}
\end{ruledtabular}
\end{table*}
\section{Summary and Conclusions}
Within the scope of this work, the nucleus $\nucl{150}{Nd}$ was investigated by means of a $\ppg$ experiment for the first time and 17 negative-parity states were observed. In particular, the branchings of the first four members for the $\kp=0^-$ octupole band were observed. Most of the remaining negative-parity states had $\kp=0^-$ character and dominantly spins with $J=3$ and 5 were excited. Three of those states show strong transitions to the $\beta$-vibrational band and can be explained by a strong mixing with states of the latter band. It is an open question if these states exist due to the inner structure of $\nucl{150}{Nd}$ or if the reaction mechanism is sensitive to states of this nature.\\
The structure of \(^{150}\)Nd was discussed in terms of the IBA-{\it spdf} model. The energy levels and \(E1\) and \(E2\) transition probabilities are reasonable well described by the calculations. The relative values are in very good agreement with the experiment, although the absolute \(B(E2)\) values are in general underestimated by the calculations. Especially for the higher spin states, the finite boson number effect plays an important role in lowering the predictions. The \(E1\) transition probabilities are also well reproduced by the calculations, especially for the \(K^{\pi}=0^{-}\) band.

\section{Outlook}
Until the present day, in 15 rare-earth nuclei branching ratios of the band heads of both the $\kp=0^-$ and $\kp=1^-$ band were observed which however is insufficient for detailed systematics. Thus, $\ppg$ experiments are essential to provide the missing information and help to understand the inner structure of the remaining nuclei. For this reason, the design and development of a new silicon detector chamber started at the end of 2010 at Universit\"at zu K\"oln. It will be included in the HORUS spectrometer in Cologne and will provide eight silicon detectors with a total efficiency of 5\%. The distances between the detectors and the target are adjustable providing the possibility to run all silicon detectors at a maximum rate. Furthermore, $\Delta E-E$ measurements will be possible giving the opportunity to perform additional inelastic scattering experiments, e.g.\ $(\mathrm{d},\mathrm{d'}\gamma)$ and $(\mathrm{p},\mathrm{t}\gamma)$ reactions. The new setup will allow measurements which will trigger new theoretical studies of octupole related structures. 

\section{Acknowledgements}
The authors would like to thank the accelerator crew of WNSL for providing the beam and the great support during our experiment. Furthermore the help of Dr.\ K.O.\ Zell concerning target preparation is highly acknowledged. CK and ME are members of the Bonn-Cologne Graduate School of Physics and Astronomy. This project has been supported by Deutsche Forschungsgemeinschaft (contracts ZI 510/4-1 and SFB 634) and U.S. DOE grant DE-FG02-01ER40609.

\bibliography{paper}

\end{document}